# Enhancing Phenotype Recognition in Clinical Notes Using Large Language Models: PhenoBCBERT and PhenoGPT

Jingye Yang[1,5], Cong Liu[2], Wendy Deng[1], Da Wu[1], Chunhua Weng[2],
Yunyun Zhou[1,3,*], Kai Wang[1,4,**]

[1] Raymond G. Perelman Center for Cellular and Molecular Therapeutics, Children's Hospital of Philadelphia, Philadelphia, PA 19104, USA

[2] Department of Biomedical Informatics, Columbia University, New York, NY 10032, USA

[3] Biostatistics and Bioinformatics facility, Fox Chase Cancer Center, Philadelphia, PA 19111, USA

[4] Department of Pathology and Laboratory Medicine, University of Pennsylvania, Philadelphia, PA 19104, USA

[5] Department of Mathematics, University of Pennsylvania, Philadelphia, PA 19104, USA

*: Corresponding author. Email: yunyun.zhou@fccc.edu

**: Corresponding author and lead contact. Email: wangk@chop.edu

**SUMMARY**

To enhance phenotype recognition in clinical notes of genetic diseases, we developed two models - PhenoBCBERT and PhenoGPT - for expanding the vocabularies of Human Phenotype Ontology (HPO) terms. While HPO offers a standardized vocabulary for phenotypes, existing tools often fail to capture the full scope of phenotypes, due to limitations from traditional heuristic or rule-based approaches. Our models leverage large language models (LLMs) to automate the detection of phenotype terms, including those not in the current HPO. We compared these models to PhenoTagger, another HPO recognition tool, and found that our models identify a wider range of phenotype concepts, including previously uncharacterized ones. Our models also showed strong performance in case studies on biomedical literature. We evaluated the strengths and weaknesses of BERT-based and GPT-based models in aspects such as architecture and accuracy. Overall, our models enhance automated phenotype detection from clinical texts, improving downstream analyses on human diseases.

**Keywords:**

**INTRODUCTION**

Rare diseases affect 30 million people in the USA and more than 300–400 million worldwide, often causing chronic illness, disability, and premature death [1-3]. Phenotype-driven approaches are increasingly used to facilitate the genetic diagnosis of rare diseases [1,4,5]. For example, a number of computational methods have been developed to facilitate phenotype-based prioritization of disease variants and genes [6-12], and some methods also enable the prediction of Mendelian diseases directly from phenotype information [13-15]. To facilitate computational phenotype analysis, the Human Phenotype Ontology (HPO) was established, which provides a standardized vocabulary to describe phenotypic abnormalities in human diseases [16]. The current release of HPO (June 2022) covers 13,000 terms and over 156,000 annotations on hereditary diseases. However, the curation of the HPO largely relies on experts' input, which may not be comprehensive enough to capture all the clinical terms, such as the age-specific neurodevelopmental and neuropsychiatric phenotypes. For instance, our previous study on children with autism spectrum disorders (ASD) [17] developed a natural language processing (NLP) pipeline and identified more ASD terms than those documented in the HPO (i.e., 3,000 terms linked to ~2,000 unique Unified Medical Language System (UMLS) concepts), representing one of the largest ASD terminology available to date. This is an illustration that NLP techniques can facilitate human reviewers in building catalogs of phenotype terms that are not previously documented for specific diseases. Therefore, there is a strong need to develop more efficient NLP methods to extract novel phenotypic concepts that may not be well covered in HPO.

Automated phenotype concept recognition from unstructured biomedical text is a type of named entity recognition (NER) task and remains a challenging problem in the biomedical NLP field. As shown in **Table 1**, we summarized three major types of methods to tackle this problem: (1) rule-based (string matching, dictionary-based, statistical model, etc.) algorithms, (2) machine learning algorithms, including recently developed deep learning methods, and (3) hybrid models combining both approaches. The first generation of tools for clinical concept recognition were either dictionary-based or rule-based approaches, such as MetaMap [18], NCBO annotator [19], ClinPhen [20], and the Aho-Corasick algorithm used in Doc2HPO [21]. MetaMap is a program developed at the National Library of Medicine (NLM) to map biomedical texts to the Metathesaurus based on a knowledge intensive approach. NCBO Annotator first creates direct annotations from raw text based on syntactic concept recognition according to a dictionary that uses terms (concept names and synonyms) from both UMLS and NCBO BioPortal ontologies, and then different components expand the first set of annotations using the knowledge represented in one or more ontologies. ClinPhen uses sequential analytic procedures with a rule-based NLP system to decide which phenotypes correspond to true mentions and which are false positives. CLAMP (Clinical Language Annotation, Modeling, and Processing) is a clinical NLP toolkit that provides not only state-of-the-art NLP components, but also a user-friendly graphic user interface that can help users quickly build customized NLP pipelines including phenotype recognition tasks [22]. In recent years, researchers began to adopt machine-learning models, including deep-learning models, due to their high accuracy and independence of hand-crafted features [23]. Machine learning based approaches, such as conditional random field (CRF) [24] and support vector machine (SVM) [25], were developed to advance the novel concept discovery; but more recently, convolutional neural network (CNN), recurrent neural networks (RNN) [26], and transformer are increasingly used. For example, the neural concept recognizer (NCR) uses the CNN to encode input phrases and then rank medical concepts based on the similarity in that vector space [27]. Transformer allows for parallelization and makes it possible to train on a much larger corpus and recognizes the long-range relationship in a sentence via its attention mechanism [28]. Transformer-based models are particularly useful for phenotype recognition since the contextual information can be very useful in determining whether a set of phrases represents a description of a particular phenotype. Pre-trained language model BERT (Bidirectional Encoder Representations from Transformers) [29] is the most well-known transformer-based model, using general corpora as the training set. The $BERT_{BASE}$ model has 12 encoders with 12 bidirectional self-attention heads, and the $BERT_{LARGE}$ model has 24 encoders with 16 bidirectional self-attention heads. Both models are pre-trained from unlabeled data extracted from the BooksCorpus with 800M words and English Wikipedia with 2,500M words.

Several studies have explored the utility of BERT-based approaches in the clinical and biomedical domains for the concept recognition task. In the biomedical domain, BioBERT [30] continued to train a model on ~18 billion words from PubMed research papers using BERT pre-trained model as the base model. BioBERT outperforms BERT in a variety of biomedical text mining tasks, which suggests that continued training on domain specific corpora improves the performance. To make the BERT model more useful in the clinical field, ClinicalBERT [31] (updated now as Bio+Clinical BERT on Huggingface model cards) started from BioBERT's pre-trained model and continued to train on 3 million notes from the MIMIC III corpora [32]. Similarly, BERN [33] used BioBERT as the backbone model to recognize known, novel, and multiple types of biomedical entities, such as gene, disease, drugs; later, the same team developed an improved version BERN2 [34] for NER using bio-lm pre-trained model. Note that these two methods focused on fine-tuning the BERT-based pre-train model for the discovery of multiple types of biomedical entities. PhenoBERT is a combined deep learning method for automated recognition of HPO; it introduces a two-levels CNN module consisting of a series of CNN models organized in two levels [35]. Furthermore, PhenoTagger was developed for HPO recognition using both dictionary-based and BERT-based model [36], and is currently one of the most reliable methods for automated HPO extraction from biomedical texts. We note that PhenoTagger can be improved in several aspects. First, PhenoTagger relies on n-gram phrase classification, so it can be difficult to differentiate different (context-dependent) concepts with the same texts or the same concept with different expressions; second, it is difficult to recognize the misspelling or low-frequency phenotypes in the samples; third, negation is not supported yet the negated instances can be important in diagnosing diseases.

As a transformer encoder-based model,  BERT has shown excellent performance on a wide range of tasks in biomedical NLP applications, including named entity recognition (described above), text classification (such as disease prediction [37]), and relation extraction (such as chemical-protein relation extraction [38]). However, little work has been done to evaluate transformer decoder-based models, e.g., GPT-type models, in biomedical NLP tasks. In our view, this is partly due to the following reasons:

1. The BERT structure was pre-trained on a large corpus so that it can be fine-tuned for a wide range of specific NLP tasks. In contrast, GPT was initially designed primarily for generating text such as in chatbots that fits the desired output format.
2. BERT was released by Google in 2018 and quickly became widely available to the research community. Although GPT was initially released by OpenAI in 2018, it is less well-known and it took a few years to improve until ChatGPT becomes widely successful. Moreover, GPT-based models were initially only available for research purposes, with limited access to commercial users.

3. Recent versions of GPT models are typically much larger and require significantly more computational resources compared to BERT models, which make them less accessible for researchers and practitioners with limited resources and budgets.

Nevertheless, recent advances in the ability to scale Large Language Models (LLMs) have resulted in top-notch performance across various NLP tasks [39]. The ability to scale LLMs to hundreds of billions of parameters has unlocked additional capabilities such as in-context few-shot learning, making it possible for LLMs to perform well on tasks trained on only a handful of examples [40]. Chain-of-Thought (CoT) prompting has also showcased the robust reasoning ability of LLMs across a wide variety of tasks, even in the absence of few-shot examples [41]. Additionally, Huang et al. have demonstrated that LLMs are capable of self-improving with only unlabeled datasets [42].

Current phenotype recognition models primarily rely on the HPO dictionary for training or prediction, limiting their ability to capture all the important phenotypic features, especially for those not well represented by HPO terms. Therefore, it will be of great scientific and practical importance to develop an accurate phenotype concept recognition model to extract all available phenotype information from clinical notes. In this study, we proposed two transformer-based models for phenotype concept recognitions: PhenoBCBERT (BERT-based) and PhenoGPT (GPT-based), and compared their performance to other existing models. We initially trained a phenotype recognition model on top of Bio+ClinicalBERT for rare disease-specific NLP text mining tasks, allowing the recognition of terms outside of the standard HPO vocabulary. Next, we implemented several GPT-based phenotype recognition models to supplement BERT-based models. Compared to existing tools such as PhenoTagger, our PhenoBCBERT can accurately infer essential phenotypic features from given contexts, despite the occurrence of non-HPO phenotypes, misspellings, and lexical dissimilarities with the original training data. Meanwhile, PhenoGPT can achieve comparable results with PhenoBCBERT with significantly fewer fine-tuning data. Therefore, both of our Transformer-based models are robust and can complement each other to achieve improved performance in concept recognition of HPO or non-HPO phenotypes.

## RESULTS

### Summary

In the sections below, we first separately described our evaluation of the performance of BERT-based (PhenoBCBERT) and GPT-based (PhenoGPT) models with comparisons to existing approaches. We then performed a comparative analysis of BERT-based and GPT-based models and discussed the relative merits of each model's architecture. Finally, we demonstrated real-world applications of these methods on publicly available clinical notes in the published genetics literature. The workflow design of the project is shown in **Figure 1**. The model architecture of PhenoBCBERT and PhenoGPT are illustrated in **Figure 2**. The tokenization techniques used in two models are described in **Table 2**.

### Evaluation of PhenoBCBERT

For the evaluation of PhenoBCBERT, our initial dataset contains 3400 automatically labeled clinical notes and 460 hand-labeled clinical notes exclusively from the CHOP database, among which 200 automatically labeled ones were randomly selected for testing and the remaining ones were used for training.

*Comparing PhenoBCBERT and PhenoTagger*

We have on average ~79% overlap of concepts identified from our PhenoBCBERT model and PhenoTagger, among the 200 test notes. By examining the test notes where >50% of entities were found in PhenoTagger but not found in PhenoBCBERT, we inferred the following two scenarios with specific examples:

1. PhenoTagger's results may contain repeated/nested phenotype mentions, yet they are skipped by our model. For instance, as shown in the clinical note 4 of **Figure 3,** PhenoTagger infers both "autism" and "autism spectrum disorder" while our PhenoBCBERT only infers "autism spectrum disorder".
2. PhenoTagger's results may contain non-phenotype entities (false positives) from human review. For instance, as shown in the clinical note 2 of **Figure 3,** PhenoTagger includes "contact the Roberts", while PhenoBCBERT does not.

There are several possibilities to explain the observed results. PhenoTagger, like other rule-based or hybrid model [35,43], utilizes localized information to facilitate predictions. Specifically, it will segment a sentence into small groups of n-grams (n = 2~10), and then feed into a deep-learning model (e.g., BioBERT) for classification. This output will be combined with dictionary-based inference for the final prediction. Since segments of local information and dictionary-based matching rely on the restricted meaning and patterns of short segments of tokens/words, it renders this method sensitive to the similarity of strings between input data and phenotype entities, which in turn leads to over-matching of the same tokens regardless of its meaning and its occurrence.

In contrast, our PhenoBCBERT takes the complete long-stretch of meaningful sentences or paragraphs as the input, passes into a fine-tuned deep-learning model, and then post-processes its output with minimal intervention (resemble subwords, etc.). Hence it will make a positive prediction only if the surrounding tokens grant it semantic meanings that resemble phenotype mentioning. Meanwhile, our infused hand-labeled data is continuously correcting the model from aligning to any wrong classifications automatically generated by PhenoTagger.

We further examined the cases where entities were identified in PhenoBCBERT but not found in PhenoTagger. We illustrated a few case studies in **Figure 3** as examples. In most cases, entities either do not have strong lexical similarity to a standard HPO term, or are written in a non-standard format (abbreviations, uncommon synonym, misspellings, etc.). For instance, in the clinical note 7 of **Figure 3,** our PhenoBCBERT successfully detected "difficulty urinating", which corresponds to "urinary retention" in the standard HPO format (HP:0000016). PhenoTagger uses the standard HPO dictionary as the training data, which includes limited amounts of standardized phenotypic abnormalities encountered in human diseases. Therefore, it is not surprising that our model can reveal many phenotype entities that are not documented in any standard dictionaries. We acknowledge that our model may also generate false positives, such as 'fragile' in Note 3. With the help of this pipeline, we will match newly found phenotype entities with higher-level HPO terms and it may help expand phenotype catalogs of rare diseases.

In addition, we found that PhenoBCBERT is more robust to typos and misspelling of words in the sentence. After removing, or replacing, random letters and/or words of phenotype entities in the testing sentence, PhenoBCBERT still recognize the compromised phenotype entities. For example, by replacing "palmar telangiectasia" with "pal telrngiectasia", PhenoTagger cannot recognize this phenotype while our model will recognize it precisely. In addition, our model can recognize abbreviations or short notation by doctors, which are not standard HPO terms. For example, 'severe IUGR' (Note 5) is not recognized by PhenoTagger.

*Evaluation of PhenoGPT*

As shown in **Table 3 and 4**, even though PhenoGPT used a significantly smaller volume of a public dataset in comparison to the in-house training data used by the PhenoBCBERT model, PhenoGPT has demonstrated highly competitive results, with an impressive accuracy of 0.857 and the top F1 score (GPT-J based) when evaluated on the BiolarkGSC+ validation dataset. In the context of prompt-based learning, we have implemented a one-shot learning strategy and through this approach, we found that the overall size of the model plays a critical role in determining its effectiveness and overall performance. GPT-J is the smallest model with 6 billion parameters; GPT-3 (davinci) is the most capable GPT-3 model with 175 billion parameters[40]; GPT-3.5 (gpt-3.5-turbo) is one of the most advanced GPT model built on GPT-3 and involves reinforcement learning with human feedback (RLHF) [44], optimized for chat. The outputs generated by GPT-J would produce incorrect phenotype entities along with arbitrary HPO IDs. In comparison, the GPT-3.5 model is capable of generating more than 80% of accurate phenotype entities, albeit with slightly compromised HPO IDs.

After the prompt-based learning, three models have drastically different performance as shown in the "GPT comparison 1" panel of **Figure 4**. The GPT-J has very low recall and makes up fake HPO IDs, whereas both GPT-3 and GPT-3.5 have reasonably good performance on entity extraction. On the other hand, GPT-3.5 is capable of performing entity normalizations by accurately assigning the appropriate HPO ID to corresponding entities. For instance, it can correctly associate "talipes equinovarus" with HP_0001762 and "foot deformities" with HP_0001760, likely due to its exposure to public phenotypic datasets during the pretraining process. Conversely, GPT-3 appears incapable of achieving such entity normalization without undergoing the fine-tuning process. However, after fine-tuning, GPT-3 is capable of successfully performing entity normalization, as discussed below.

The prediction results after fine-tuning are illustrated in the GPT comparison 2-6 of **Figure 4**. We found that the performance of the model is significantly influenced by the size of the fine-tuning dataset. Yet, both the closed-source and open-source models can yield comparably favorable outcomes. As shown in the "GPT comparison 2" panel (**Figure 4**), the performance of the GPT-3 based model fine-tuned on 28 instances is even worse than the prompt-based learning. Conversely, other model finetuned on 200 instances has considerably better results. It has fewer false positive predictions and more accurately normalized HPO IDs compared to prompt-based results. For example, the fine-tuned GPT-3 (w/ 200 instances) can successfully extract "distal arthrogryposis (HP: 0005684)" and "foot deformities (HP: 0001760)", which were in the results of prompt-based GPT-3 prediction but with incorrect HPO IDs (HP: 0001812 and HP: 0000827, respectively); the fine-tuned model also excluded false outcomes in the prompt-based learning like "strict diagnostic criteria", "incomplete ascertainment", "range of phenotypes", etc. It is not surprising that the fine-tuned model will be more reliable and accurate on the phenotypic entity recognition from prompt-based learning, since the model will compute gradients and make updates on all of its parameters to fit the fine-tuning dataset. In addition, we searched through the 200 notes in the fine-tuning dataset and found that all phenotypic entities with the correct HPO ID normalization have appeared at least once in the fine-tuning dataset, which explains the reason why the fine-tuned model can successfully normalize those phenotypes.

As illustrated in the GPT "comparison 3" panel (**Figure 4**), both the closed-source and open-source models demonstrate strong ability to identify phenotype entities accurately. For example, in the case of the previous 9 positive phenotypic entities, both the GPT-J-based and Falcon-based open-source models were able to successfully identify all of them. The LLaMA-based open-source model, however, missed one phenotype, "arthrogryposis", but it still correctly classified it under the more comprehensive HPO term, distal arthrogryposis (HP: 0005684). It is important to note that both LLaMA and GPT-J generated an additional false positive each; LLaMA misidentified "incomplete ascertainment", and GPT-J misidentified "variable". On the other hand, the Falcon model demonstrated an impeccable match for all phenotype entities.

We further evaluated the performance of our open-source models across various clinical abstracts, as showcased in GPT comparisons 4-6 (**Figure 4**). We found that these three PhenoGPT open-source models demonstrate consistency and strong performance. In rare instances, there are discrepancies in results, as seen in GPT comparison 6. The GPT-J-based model predicted two additional true positives, but also reported a false positive, "miscarriage". Both LLaMA and Falcon predicted one additional true positive each, namely, "coxa-epiphysiolysis" and "obesity".

In conclusion, the results presented above emphasize the critical role played by both the size and quality of the pretraining data in determining the final performance of the model. Furthermore, it is highly recommended to initially fine-tune the large base GPT model on domain-specific datasets before utilizing it for inference, as this process enhances the model's ability to perform accurate entity normalization. When it comes to open-source versus closed-source models, the choice largely depends on an individual's specific requirements. Open-source models offer greater flexibility, allowing for customization and the use of various training strategies such as LoRA, quantization, and others. On the other hand, closed-source models depend on third-party services and are typically easier to implement through API calls. Both types of models demonstrate commendable performance.

### *Comparison between the BERT-based model and GPT-based model*

#### <u>Performance on phenotype entity recognition</u>

Both PhenoGPT and PhenoBCBERT are capable of extracting phenotypic information from unstructured raw clinical data. They have comparably high precision, recall and accuracy on the public validation dataset. As shown in the **Table 3 and 4**, PhenoGPT has the best Recall and F1 score for certain datasets while PhenoBCBERT has relatively lower scores. Since we formulated NER as a causal language model task for GPT model's fine-tuning, it requires fewer efforts to post-process its outcomes (e.g., HPO ID normalization). In contrast, PhenoBCBERT needs extra modules to normalize its phenotype entity predictions (e.g., Sent2Vec or SVM). The precision scores of our models are not superior, which is partially due to the fact that our models can detect the words that are not included in the standard HPO dictionary.

We also observed the nuances in the performance of the PhenoGPT model family, which showed slightly inferior results on the ID-68 dataset compared to PhenoBCBERT. These nuances can be attributed to various factors including dataset characteristics, architectural decisions, and training approaches. Specifically, ID-68 dataset comprises real-world clinical notes focused exclusively on families with intellectual disabilities, which naturally limits the model's performance to specific areas. PhenoBCBERT excels particularly in recognizing entities related to intellectual disabilities, likely owing to the high representation of this specific phenotype in the in-house dataset. This contrasts with other larger language models, which demonstrate better overall performance, as illustrated in **Table 3**. Another point to consider is the inherent difficulty in training large GPT-based language models. These models often rely heavily on their pre-existing knowledge base for decision-making, which is precisely why such large-scale models are developed in the first place. One of our primary objectives is to fine-tune large language models in a cost-effective manner to excel at general Named Entity Recognition

(NER) tasks in the biomedical domain. Our findings indicate that both a fully fine-tuned BERT model and an efficiently fine-tuned large GPT model can outperform other general-purpose models in these tasks. This validates our approach and bolsters our confidence to continue refining our large language models for larger cohorts in future projects.

*Data efficiency for model training*

Despite the differences in model architecture and training approaches between PhenoBCBERT and PhenoGPT, both models demand accurately labeled data containing phenotypic information. Therefore, gathering and preprocessing raw clinical data for both models are equally challenging and expensive. Nonetheless, since GPT has been pretrained on a substantially larger dataset compared to BERT, it needs a relatively smaller fine-tuning dataset to attain comparable outcomes. Note that it is also possible that GPT models might have already seen the public datasets that were used in the fine-tuning step.

*Computing resource requirement and accessibility*

PhenoBCBERT has 110 million parameters and requires 1-2GB of RAM during training, making it suitable for fine-tuning with larger batch sizes. The closed-source PhenoGPT can have up to 175 billion parameters, demanding 600 GB of RAM, so fine-tuning is performed using OpenAI's API and stored in the cloud for inference. The open-source PhenoGPT models still necessitates a certain amount of GPU resources, ranging from 14-70GB, depending on the specific model and training strategy utilized. The open-source model was saved locally and sharable for public use.

In conclusion, BERT is more affordable and efficient, while GPT has excessive capabilities. The choice of which model to use depends on task requirements, and it is advisable to test different models since they each have their own merits.

### *Additional validation through case studies*

We applied both BERT-based and GPT-based models (GPT-3) on clinical notes selected from papers published by the American Journal of Human Genetics (AJHG) [45-47] to extract phenotype entities. As shown in **Figure 5**, we noticed some differences between the results of PhenoTagger and our models. We underlined PhenoTagger-specific outputs, PhenoBCBERT-specific outputs, and PhenoGPT-specific outputs in green, red and blue, respectively. We also highlighted negation detection and misspelled entities in red and yellow.

In the example of **Figure 5A**, we observed that PhenoBCBERT captured all the phenotype entities, including two essential phenotypes missed by PhenoTagger ("restricted cerebellar growth" and "problems with latching and swelling") and one negated phenotype entity ("chromosomal abnormalities", which is an "abnormal cellular phenotype"). In addition, PhenoTagger mistakenly recognized one non-phenotype entity "nuchal translucency" as "Increased nuchal translucency" (HP: 0010880). PhenoGPT had high precision and skipped the negated phenotype entity, but it could not identify "ventriculomegaly" and "small cerebellum".

In the example **Figure 5B**, PhenoTagger could not recognize phenotypes "pyelocaliceal dilatation", "dysmorphisms" and "micrognatia". In contrast, PhenoBCBERT successfully highlighted all essential phenotypes, with extra granularity ("learning difficulty in expressive language" instead of "learning difficulty"). We also noticed that PhenoBCBERT did not identify non-phenotype entities such as "vacuum extraction". For this note, PhenoGPT did not over-predict phenotype terms such as "gestational diabetes" (maternal phenotype) and "umbilical artery" (non-phenotype). Additionally, PhenoGPT identified the less obvious phenotype entity, "height and weight above the 95th centile".

In the example **Figure 5C**, PhenoTagger could recognize most of the positive phenotype entities, while both PhenoBCBERT and PhenoGPT slightly outperformed with more true positive predictions ("severe neurosensory hyopoacusia") and more accurate phenotypic description ("… with stiffness", "right-convex scoliosis", "Achilles tendon", "tendon retraction"). Our models missed one general phenotype "Pyramidal signs", but successfully captured its child term "high tendon reflexes", which contains more information at its level of taxonomic hierarchy.

Interestingly, in the example **Figure 5B** and **Figure 5C**, we noticed misspellings in the original data (these mistakes are present in the originally published AJHG manuscript): "micrognatia" should be "micrognathia", and "hyopoacusia" should be "hypoacusis", which disguised themselves from PhenoTagger yet both PhenoBCBERT and PhenoGPT could recognize them. These examples supported the robustness of our transformer models as discussed above.

The under-performance of PhenoTagger on these applications is mainly caused by two reasons, as specified in the original paper [36]: first, PhenoTagger cannot disambiguate the different concepts with the same text name; second, PhenoTagger misses some phenotype concepts that are lexically dissimilar with the concept terms in the HPO.

In conclusion, these case studies demonstrate that context information can be important and informative to reliably make predictions on phenotype recognition tasks. PhenoBCBERT and PhenoGPT can accurately infer essential phenotypic information from the given context, despite the occurrence of rare phenotypes, misspelling, and lexical dissimilarity with the original training data.

## DISCUSSION

The study used BERT-based and GPT-based models on clinical texts from pediatric patients to identify known and unknown (not documented by HPO) clinical phenotypes. We found that both PhenoBCBERT and PhenoGPT can identify new concepts with better accuracy, recall, and precision than competing models. The methods can be adapted to other biomedical domains and can identify novel entities based on context. Despite the strengths, these models have several limitations and further improvements can be made.

### *Effect of data quantity and quality*

Both PhenoBCBERT and PhenoGPT require annotated training data. The performance levels of the fine-tuned models may vary based on the quality of data collected and the labeling strategy utilized. Additionally, there is a potential for training bias, especially if physicians from the same institutions or teams tend to document notes in similar styles. For instance, physicians often repeatedly include a patient's past clinical history in a current clinical note when drafting a new note, which may lead to a bias toward specific semantic structures or recurring tokens (when the same passage occurs multiple times in the same note). Additionally, unexpected errors by doctors (such as typos, missing words, or incorrect phenotypes) can hinder the automatic labeling process from generating accurate training data. Moreover, the notes may contain redundant yet potentially privacy-leaking information (such as a patient's address or first name), which raises serious privacy and ethical concerns in the model training [48]. Although we used a Stanford-Penn MIDRC Deidentifier model for de-identifying all clinical notes before using them in the BERT training process, we are still concerned about the possibility of leaking private information from patients and therefore cannot share the model trained on in-house data. On the other hand, PhenoGPT, which can utilize both open-source and closed-source GPT models, was fine-tuned using only minimal amounts of data available to the public. Thus, we do not have the same privacy leakage concerns that BERT-based models might pose. Despite this, thanks to their extensive pre-trained knowledge base, they still demonstrate strong performance on the evaluation datasets. In our future research, we aim to incorporate more diverse datasets sourced from various healthcare systems for additional refinement/improvements of the phenotype recognition models, with appropriate IRB protocols. We will assess the degree to which these institutional datasets enhance our models' effectiveness and generalizability. Lastly, we discussed above generic concerns applicable to all deep learning models when used on real clinical data. Several different existing strategies, including strings comparison for repeated tokens, auto-correction for misspelling [49], de-identification model [50] to remove patient information, have been explored to minimize the undesirable negative impacts of raw clinical data.

### Model structure

For the BERT-based model, we used pretrained Bio+Clinical BERT as the main deep learning model, which is configured from $BERT_{base}$. The size of the model (number of attention layers, number of attention heads, dimensions of non-linear layer, etc.) is relatively small compared to recently published large language models. In this case, pretraining for a larger model with billions of parameters on EHR data may be necessary. Based on the promising result from the current study, we will pretrain a specialized large language model on de-identified clinical notes from our database for future downstream EHR-NLP tasks.

We also noticed the restrictions on the length of sentences for the language model (512 tokens for $BERT_{base}$ and $BERT_{large}$). To overcome the bottleneck, we will further adopt the down-sampling [51] or/and filtering methods [52] on top of our model. Meanwhile, we can distribute many independent models on multiple GPUs for training such that a combination of them will take care of extra-long samples [53].

With the GPT-based model, its larger input window and automatic normalization are advantageous as part of the output. However, the vast number of parameters and associated costs can make it difficult for retraining. Moreover, as a generative language model, its strengths in chat completion and next-sentence generation were not fully utilized. In fact, our fine-tuning process diminished its capacity to generate coherent human-like responses. Consequently, a more effective strategy to fully leverage the model's structure is essential.

### Model Selection and Implementation

Our PhenoBCBERT model, initialized from a BERT-based architecture, demands the fewest computational resources. However, its Named Entity Recognition (NER) training strategy necessitates additional steps for entity normalization and negation, potentially resulting in a longer overall processing time. On the other hand, while our GPT-based models do require a certain degree of computational power during training, their final versions are relatively resource-efficient and easy to deploy for inference and predictive tasks.

In comparison to PhenoTagger, both of our models can be conveniently transferred to a local machine for making predictions. Even for setups with limited GPU resources, CPU-based deployment remains a viable alternative. For users interested in custom training our models on their datasets, we have provided comprehensive guidelines and scripts in the code repository that detail how to fine-tune large language models with minimal computational expenditure.

### Future Directions

Although both our models show promising results for phenotype entity recognition, they cannot classify all of the desired phenotype entities in the context. Concretely speaking, we list the following two potential improvements in terms of our current work.

### *Phenotype entity recognition*

In the test data, we noticed that our model would recognize medication names and disease names as positive if they appear in a similar context as phenotypes. (e.g., he has **red eyes** vs. he has **sertraline**.) Also, despite post-processing, the output of BERT-based model may skip prepositions in phenotype entities (**learning difficulty** in the **expressive language**). In very few cases, both models missed out on rare phenotype entities.

Incorporating hard negative examples into the annotated dataset for fine-tuning could enhance phenotype entity recognition performance. Additionally, employing multi-modal networks to boost natural language understanding may be another potential approach. These strategies represent potential avenues for future exploration.

### *HPO normalization and negation*

We observed that our model can encounter errors when associating phenotype entities with their corresponding HPO IDs. This issue arises as each HPO term may have multiple synonyms or typographical errors that were not present in the training data. Furthermore, PhenoBCBERT employs Sent2Vec to calculate cosine similarity between a given phenotype entity and all HPO terms, assigning the most similar HPO term to the phenotype entity. Meanwhile, PhenoGPT conducts the next token prediction based on a given phenotype entity. Neither of these approaches can effectively tackle normalization problems.

We hypothesize that employing vector embedding could enhance the HPO normalization process. In particular, we aim to leverage GPT's extensive knowledge to generate a database of HPO embeddings as high-dimensional vectors, where each vector is normalized to enable either Euclidean or cosine distance as similarity measures. Such an embedding should yield more accurate normalization outcomes and better interpretability. Lastly, we can enhance the detection of negated entities by incorporating a dedicated post-processing model based on the transformer or LSTM architectures as a part of the pipeline, rather than relying on general NLP packages such as *Negspacy*. This will be one of our focuses in future research.

In addition to the two aspects previously discussed, we have also fine-tuned the latest version of the LLaMA 2 model to investigate whether it offers any advancements in HPO normalization or notable improvements in phenotype entity recognition. Unfortunately, the most recent LLaMA 2 iteration does not yield better outcomes; its performance closely mirrors that of the original LLaMA. We have made this model accessible in our public code repository for further testing. Going forward, we plan to delve into the utility of large-scale generative language models for a range of biomedical NLP tasks, such as entity recognition, relationship extraction, and text classification, in a cost-effective manner. Given the strength of reinforcement learning, we also hope to develop such a model capable of adapting to changes in the environment and can continue to learn over time (for example, personalized biomedical NLP pipelines). Meanwhile, scaled models will have compromised interpretability and credential issues, and carefully addressing these problems is another direction for future work.

## EXPERIMENTAL PROCEDURES

### Resource availability

#### *Lead contact*

Further information and requests for data should be directed to and will be fulfilled by the lead contact, Dr. Kai Wang (wangk@chop.edu).

*Materials availability*

This study did not generate new unique materials.

*Data and code availability*

The various GPT-based models and example notebook scripts are available to download at https://github.com/WGLab/PhenoGPT. All original code has been deposited at DOI: 10.5281/zenodo.8346470 and is publicly available as of the date of publication. Any additional information required to reanalyze the publicly available data reported in this paper is available from the lead contact upon request.

### Summary

PhenoBCBERT utilized encoder-based BERT model by fine-tuning the pretrained Bio+Clinical BERT model for concept recognition. PhenoGPT utilized decoder-based GPT model (GPT-J, Falcon [54], LLaMA [55], GPT-3 [40], GPT-3.5 [44]) and can be trained by either prompt-based strategy or fine-tuning strategy. The framework of the project design is shown in **Figure 1**. The workflow of both PhenoBCBERT and PhenoGPT is illustrated in **Figure 2.** All the computational models were built using the PyTorch module and HuggingFace [56] interface.

### Data preparation

This study was approved by the institutional review board of the Children Hospital of Philadelphia (CHOP). Our in-house dataset consists of clinical notes of rare disease patients seen at CHOP. We used 2252 ICD10 code [57] (that can map to the rare disease in the Orphanet Database) to query rare disease patients. Among these, we sub-sampled more than 4500 clinical notes regarding rare diseases, each of which contains potential phenotypic concepts that have been recognized by physicians. Although these potential phenotype entities are not complete, they will serve as the foundational phenotype information to assist in locating clinical abstracts and incorporating additional related phenotype entities using the mixed labeling strategy outlined in the following section. Since patients might have multiple visits to the hospital due to different reasons, we only retained disease progression summary notes that map to these rare-disease specific ICD10 code. We further excluded low-quality notes defined as patients with less than 4 visits or notes with less than 1000 words. Each progression summary note contains full counseling history, including medical history, surgical history, testing and imaging, developmental history, physical history, family history, genetic counseling summary, etc. In the end, we obtained 3860 high-quality clinical abstracts by using associated fundamental phenotypic concepts as the query to extract the counseling summary, removing extraneous information, and then truncating each summary to a maximum of 2400 characters (approximately 500 tokens). This was necessary as we are utilizing a BERT-based model that can only accommodate up to 512 tokens per sample. After truncation, the 3860 clinical abstracts contained over 14,000 individual trainable sentences. Due to confidentiality concerns, we used this data locally only for PhenoBCBERT.

Our publicly available dataset consists of BiolarkGSC+ [58] and ID-68 [35]. BiolarkGSC+ is an updated version of Bio-Lark gold-standard corpus (GSC) dataset [59] with 228 de-identified clinical notes abstracts and the corresponding HPO terms. The ID-68 dataset includes 68 de-identified clinical notes from families with intellectual disabilities [60] and with HPO terms annotated in the same way as in the GSC+ dataset [35]. These public datasets were used mainly for fine-tuning, validating GPT-based models, and comparison between various rule-based and deep-learning models. Furthermore, in our case studies, we also evaluated performance of different methods on three distinct clinical abstracts of published biomedical literature in the American Journal of Human Genetics (AJHG) [45-47] as the independent third-party data source. Specifics will be covered in the extra case study section.

### Tokenization

In the PhenoBCBERT, we used WordPiece tokenization and three different embeddings, consisting of token, position, and sentence embedding, to represent the input information [29]. Most meaningful words are kept and the other words are tokenized into pieces. An uncommon word can be split into more than one tokens, and two sharps (##) are added in front of the tokens. For example, "arthritis" is tokenized into "art", "##hr", "##itis". This means the word "arthritis" was less common than other words when training the WordPiece representation. This technique helps us to tokenize all possible words in the literature regardless of whether they occurred before. Two special tokens [CLS] and [SEP] were used to mark the start and the end of a sentence. We did not consider the order of sentences in our model, so the sentence embedding will default to all 0's.

In the PhenoGPT, we used GPT's fast Byte Pair Encoding (BPE), which can handle out-of-vocabulary (OOV) words in a similar fashion as WordPiece tokenization with additional SentencePiece [61] mechanism. Since GPT is a generative decoder model, we do not need sentence embeddings. **Table 2** summarized the tokenization techniques used in two models.

### Labeling and Training strategy

To train PhenoBCBERT and PhenoGPT models for phenotype entity recognition, different labeling strategies were required due to the fundamental differences in their model structures.

#### *PhenoBCBERT*

We labeled our BERT-based model PhenoBCBERT following the standard NER (name entity recognition) task practice. We utilized PhenoTagger automatic labeling in our training process because (1) it is a hybrid model combining both deep-learning-based results and rule-based classification, therefore making it a good representative of two major classification methods; and (2) it outperforms other state-of-the-art methods for phenotype concept recognitions, like Doc2hpo [21], OBO [62] and NCR [27].

To avoid duplicating PhenoTagger's results, we will use data augmentation, as in various NLP tasks [63-66], to improve our training dataset by infusing the 3400 automatically labeled clinical notes with 460 hand-labeled clinical notes from our in-house dataset. We name this labeling strategy with data augmentation as mixed-supervision, in contrast to a fully-supervised dataset with manually labeled clinical notes or a dataset with labels generated completely by third party NER machines.

Finally, we located starting and ending positions of each phenotype entity from PhenoTagger's output file in the original input sentence and matched them to the tokenized subwords for labeling as shown in the table. Instead of the IOB labeling strategy which distinguishes beginning and inside tokens, we simply adopted a binary inside-outside labeling strategy for efficiency and labeled all positive sub-word tokens as 1. The final labeling is illustrated in **Table 2**. The starting and ending position are manually annotated in 460 hand-labeled clinical notes.

#### *PhenoGPT*

Instead of training a GPT model for name entity recognition, we labeled the training data to comply with the model's nature as a generative decoder. For a given clinical abstract, we generated a text by appending phenotype entities with their associated HPO IDs to the abstract for either prompt-based learning or fine-tuning.

Comparing these two models' training strategies, although NER needs a lot of time and effort on data preprocessing (mixed labeling) and labeling, it is much easier to train than causal language modeling, as in a GPT model. Causal language modeling cannot attend to future tokens and can only make predictions on the next token, whereas the NER model can attend to a complete sentence to make a classification on the [CLS] token. In general, generative language models require a larger amount of pretraining data due to a larger amount of training parameters, which may cause unexpected behaviors [67].

### *Model initialization, fine-tuning and prompt-based learning*

Our PhenoBCBERT was initialized from the Bio+Clinical BERT [31] as the pre-trained base model. After initialization, we continued to finetune all parameters with our mixed-supervised dataset and apply an extra token classification layer for label prediction (out-of-bag or phenotype entity). Lastly, we compared our results with PhenoTagger's output to map common phenotypic entities and rely on Sent2Vec's similarity score [68] of embeddings to normalize newly detected phenotypes in a manner similar to this study[17]. We also applied Negspacy [69] to exclude negated entities in the final step.

Our PhenoGPT model was constructed upon a range of GPT models, with the specific choices influenced by factors such as model availability and size, as we aimed to ensure ease of reproducibility for users. For the open-source models, we utilized GPT-J-6B, Falcon-7B, and LLaMA-7B as the initial models. In contrast, for the closed-source models, we opted to start with the GPT-3 model. Both versions were subsequently fine-tuned using the public BiolarkGSC+ dataset. It is also worth mentioning that we did not employ any proprietary data in this task to avoid the potential risk of disclosing confidential information, particularly with the closed-source model. We used cross-entropy loss to penalize the causal language model for producing incorrect next tokens, i.e., wrong phenotype entities or HPO IDs. Since a large language model like GPT is capable of prompt-based few-shot learning, we also tested GPT-J, GPT-3, and GPT-3.5 for their performance given only prompts. We used the prompt to guide the GPT model towards generating phenotypic features. For example, the prompt we used is " *please identify human phenotype ontology for me*". To evaluate the performance, we used the publicly available BiolarkGSC+ dataset and ID-68 dataset for training and validation, and both datasets include human labeled HPO terms and their identifiers. We divided the BiolarkGSC+ dataset into two asymmetric potions, one containing 200 instances and the other one containing 28 instances. We then proceeded to fine-tune our entire suite of PhenoGPT models on a larger dataset comprising 200 instances. Furthermore, to assess the impact of training sample sizes, we also fine-tuned two PhenoGPT models based on GPT-3: one was trained using a dataset of 200 instances and validated on a set of 28 instances over 12 epochs, while the other was trained on a smaller set of 28 instances and validated on the larger 200-instance set over 32 epochs. Note that these two models were also referred to as GPT-3 (200) and GPT-3 (28) in the **Figure 4**.

In the deployment of open-source PhenoGPT models (GPT-J, Falcon, LLaMA), we used 4-bits and 8-bits [70] model quantization strategy to reduce our model size to only 1/4$^{th}$ without compromising the performance. Following this, the language model parameters are set as fixed and a comparatively small quantity of trainable parameters are incorporated into the model via Low-Rank Adapters (LoRA) [71]. During the fine-tuning process, the optimizer propagates gradients through the static 4-bit quantized pre-trained language model into these Low-Rank Adapters. It is important to note that only the LoRA layers undergo updates during the training phase. This training strategy, known as QLoRA [72], is implemented across all of our open-source PhenoGPT models.

## ACKNOWLEDGEMENTS

We thank Dr. Li Fang for providing insightful suggestions and guidance. This study is in part supported by NIH grant LM012895, HG012655, HG013031, Federal work-study program, Penn DDDI fellowship program and CHOP Research Institute. We thank technical support from the IDDRC Biostatistics and Data Science core (HD105354), and CHOP Information Services for support on GPU computing.

## AUTHOR CONTRIBUTIONS

Conceptualization, J.Y., C.W., Y.Z. and K.W.; Methodology, J.Y., Y. Z., and K.W.; Software, J.Y. Investigation, J.Y., W.D., Y.Z., and K.W.; Data Curation, J.Y., W.D., and Y.Z.; Writing – Original Draft, J.Y., D.W., Y.Z., and K.W.; Writing –Review & Editing, J.Y., C.L., W.D., D.W., Y.Z. and K.W.; Resources, K.W.; Supervision, K.W.; Funding Acquisition, K.W.

## DECLARATION OF INTERESTS

The authors declare no competing interests.

**Legends of Figures**

Figure 1. Illustration of the workflow of the project.

Figure 2. Illustration of the BERT-based and GPT-based model used in the current study for a sentence with phenotype mention. (A) Conversion of input sequence into a combination of three embeddings (sentence embedding is treated as zero in the current study). (B) The pre-training and fine-tuning strategy for PhenoBCBERT. (C) The pre-training and fine-tuning strategy for PhenoGPT.

Figure 3. Examples of phenotype terms from 8 clinical notes recognized by PhenoTagger and PhenoBCBERT. The font size is relative to the frequency of appearance.

Figure 4. Examples of phenotype terms from clinical notes recognized by different series of GPT models. GPT comparison 1: prediction results after prompt-based learning. GPT comparison 2-6: prediction results after fine-tuning. GPT-3 (N): GPT-3 fine-tuned based on N instances.

Figure 5. Case studies of the predicted phenotype entities with PhenoTagger, PhenoBCBERT and PhenoGPT (GPT-3). The negation terms and misspelled terms from the original published manuscript are highlighted.

**TABLES**

| | Rule-based | Hybrid | Deep-learning | Year |
|---|:---:|:---:|:---:|:---:|
| Metamap | ✓ | | | 2001 |
| NCBO | ✓ | | | 2009 |
| OBO | ✓ | | | 2014 |
| Doc2Hpo | ✓ | | | 2019 |
| ClinPhen | ✓ | | | 2019 |
| NCR | | ✓ | CNN (local) | 2019 |
| Phenotagger | | ✓ | BERT(local) | 2021 |
| BERN2 | | | BERT | 2022 |
| PhenoBERT | | ✓ | CNN+BERT | 2022 |

Table 1. Summary of different phenotype recognition models.

| Input | BERT Tokenize | GPT Tokenize | Labeling |
|---|---|---|---|
|  | [CLS] |  | -100 |
| he | he | he | 0 |
| has | has | _has | 0 |
| red | red | _red | 1 |
| eyes | eyes | _eyes | 1 |
| and | and | _and | 0 |
| distance | distance | _distance | 1 |
| exotropia | exo | _exo | 1 |
|  | ##tropia | tropia | 1 |
|  | [SEP] |  | -100 |

Table 2. Illustration of labeling strategies for a sentence with phenotype mentions.

|  | BiolarkGSC+ (Validation) | | |
|---|---|---|---|
| Model | Precision | Recall | F1 |
| OBO Anotator | 0.810 | 0.568 | 0.668 |
| NCBO | 0.777 | 0.521 | 0.624 |
| Doc2hpo-Ensemble | 0.754 | 0.608 | 0.673 |
| MetaMap | 0.707 | 0.599 | 0.649 |
| Clinphen | 0.590 | 0.418 | 0.489 |
| NeuralCR | 0.736 | 0.610 | 0.667 |
| PhenoTagger | 0.720 | 0.760 | 0.740 |
| PhenoRerank | **0.843** | 0.708 | 0.770 |
| PhenoBCBERT | 0.747 | 0.813 | 0.779 |
| PhenoGPT(GPT-3) | 0.827 | 0.794 | 0.810 |
| PhenoGPT(GPT-J) | 0.809 | **0.857** | **0.832** |
| PhenoGPT(Falcon) | 0.801 | 0.792 | 0.796 |
| PhenoGPT(LLaMA) | 0.828 | 0.694 | 0.755 |

Table 3. Performance comparison on the BiolarkGSC+ validation set. PhenoGPT and PhenoBCBERT were fine-tuned using the BiolarkGSC+ training set.

| | ID-68 (Validation) | | |
|:---:|:---:|:---:|:---:|
| **Model** | **Precision** | **Recall** | **F1** |
| NCBO | 0.874 | 0.660 | 0.752 |
| MetaMapLite | 0.804 | 0.591 | 0.682 |
| Doc2hpo | 0.844 | 0.575 | 0.684 |
| Clinphen | 0.749 | 0.615 | 0.675 |
| NeuralCR | 0.786 | 0.776 | 0.781 |
| PhenoTagger | 0.898 | 0.755 | 0.820 |
| PhenoBERT | **0.943** | 0.781 | 0.854 |
| PhenoBCBERT | 0.912 | 0.923 | **0.872** |
| PhenoGPT(GPT-3) | 0.818 | 0.814 | 0.816 |
| PhenoGPT(GPT-J) | 0.723 | 0.758 | 0.740 |
| PhenoGPT(Falcon) | 0.738 | 0.881 | 0.803 |
| PhenoGPT(LLaMA) | 0.719 | **0.926** | 0.809 |

Table 4. Performance comparison on the ID-68 dataset. PhenoGPT and PhenoBCBERT were fine-tuned using BiolarkGSC+ training set.

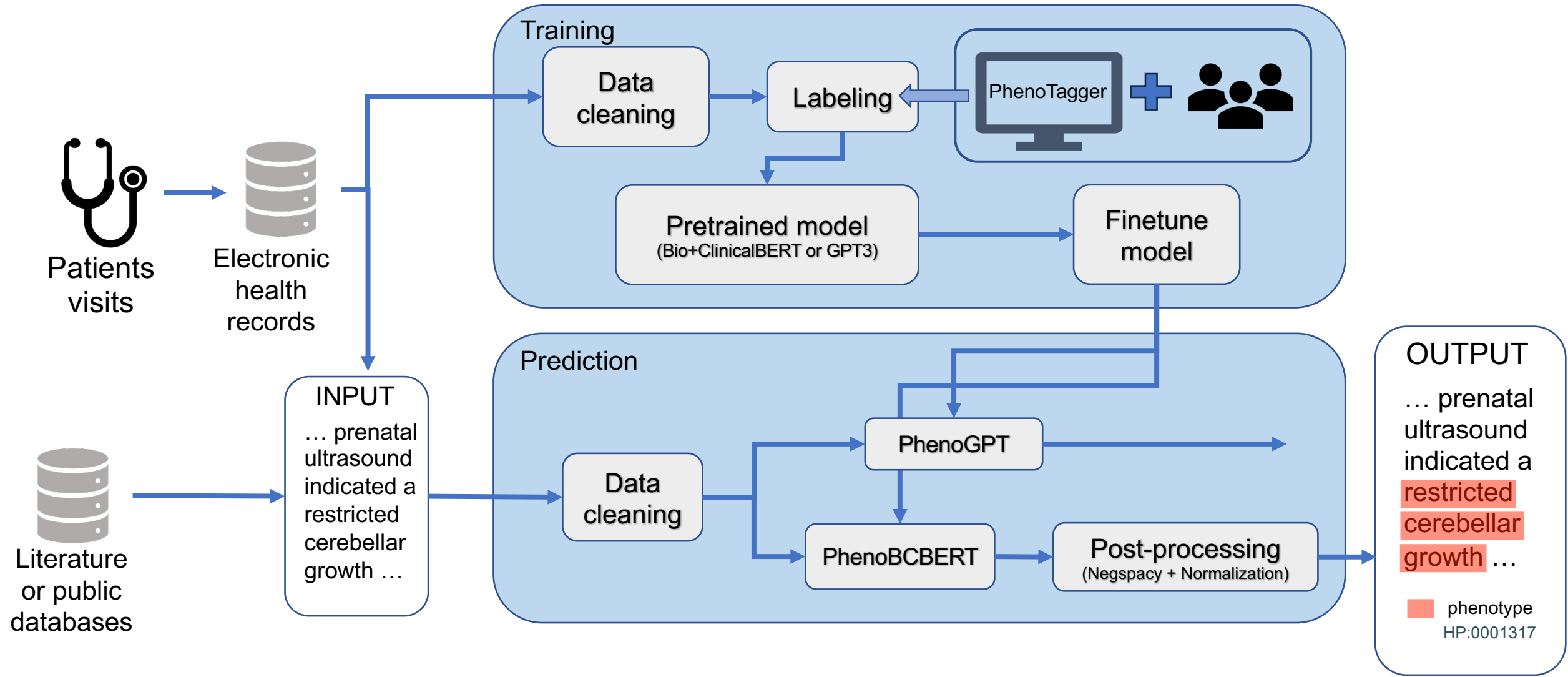

**A**

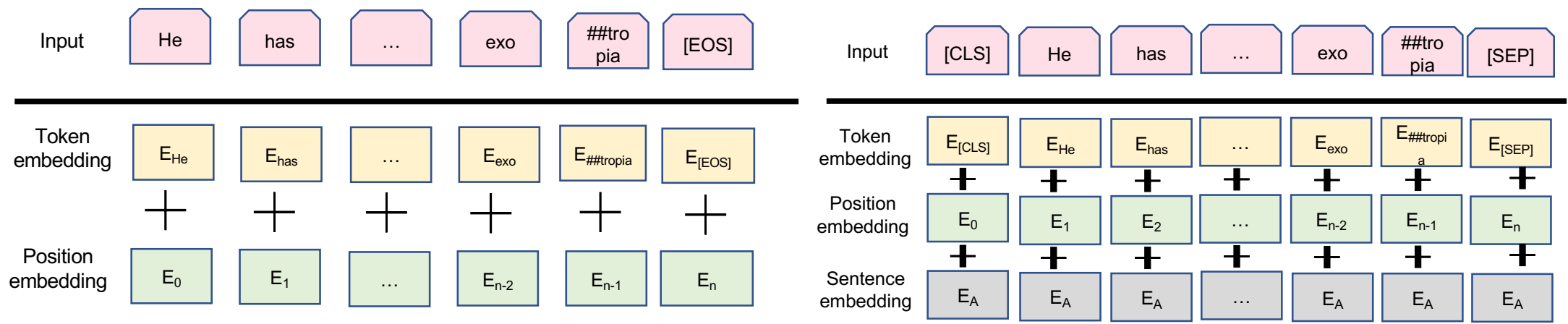

**B**

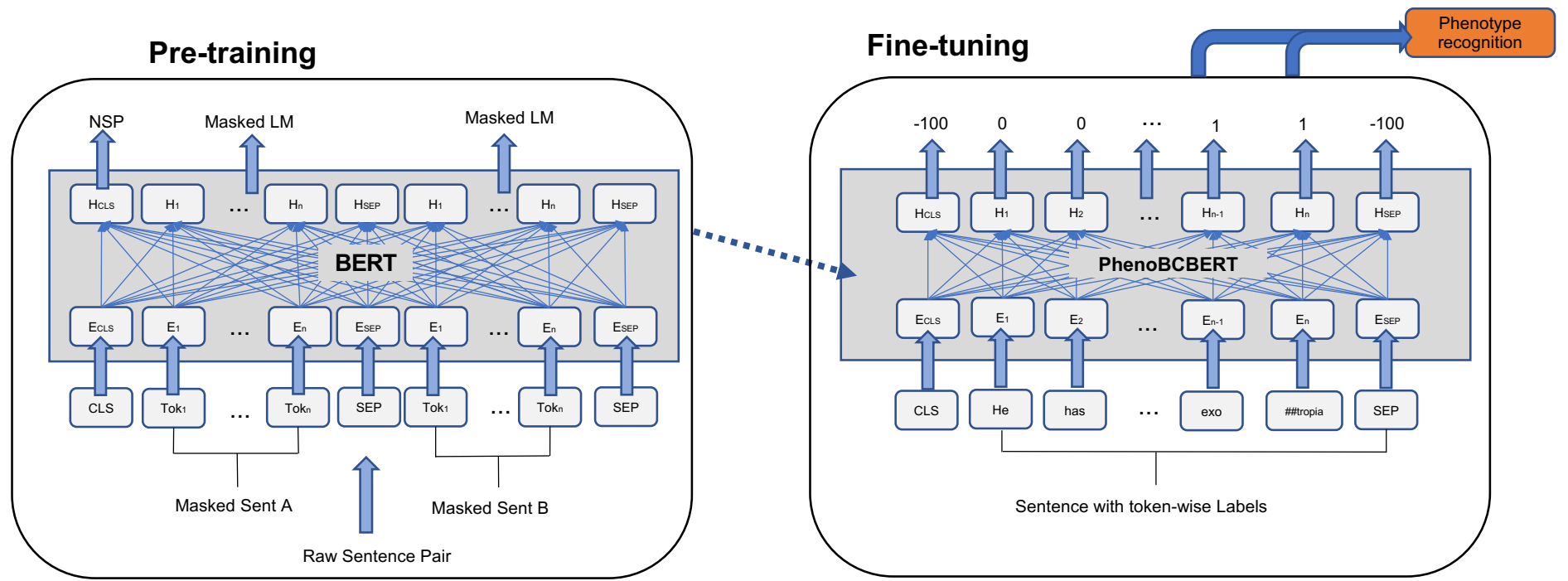

**C**

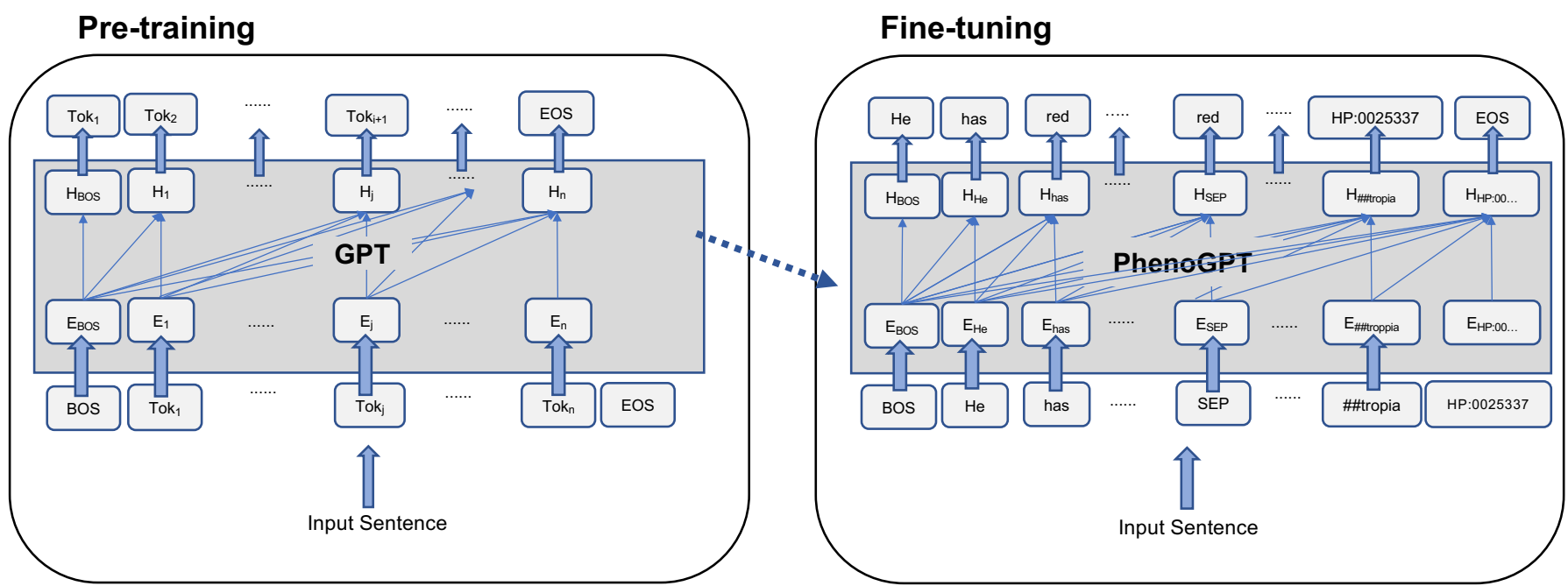

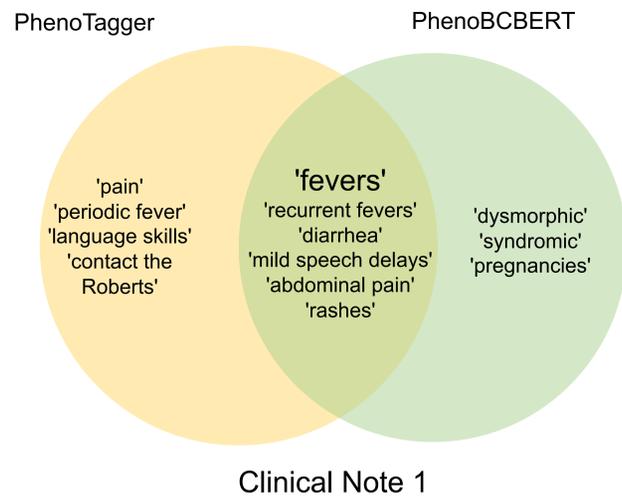

Clinical Note 1

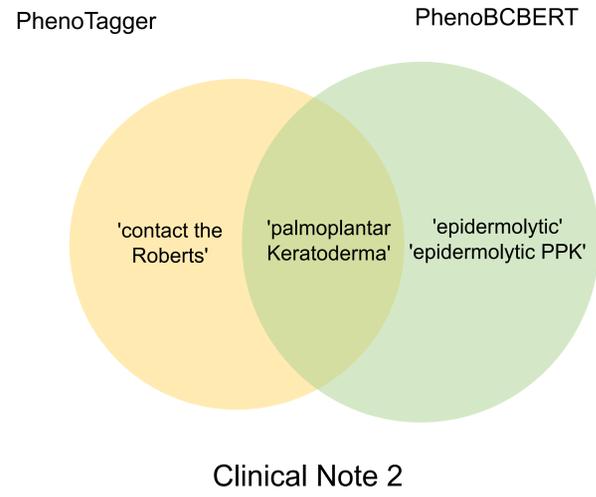

Clinical Note 2

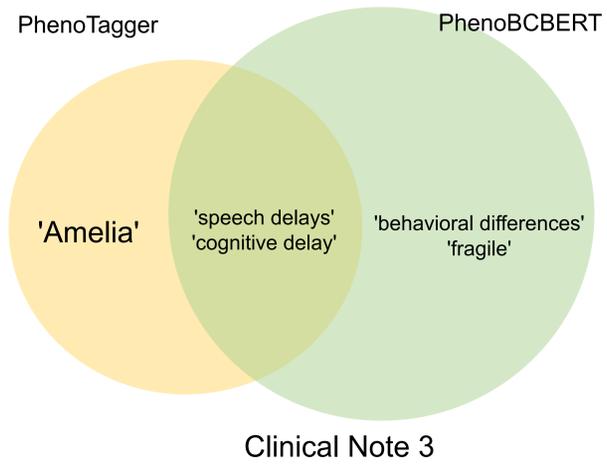

Clinical Note 3

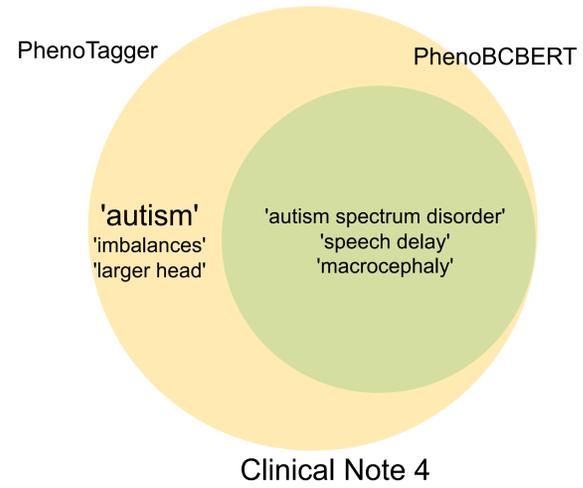

Clinical Note 4

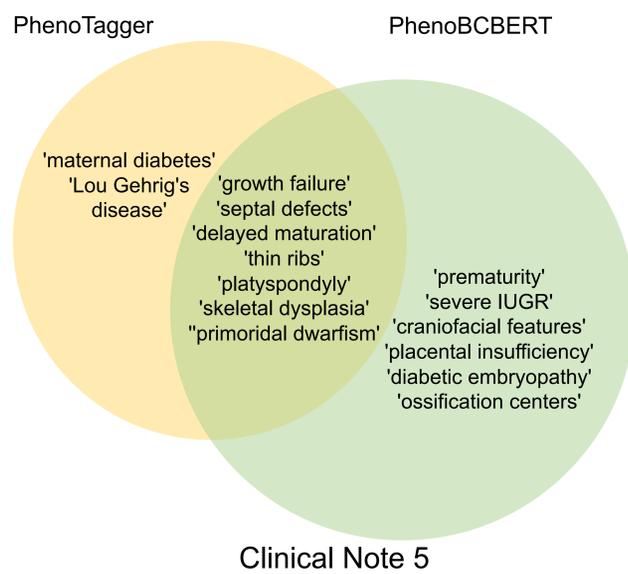

Clinical Note 5

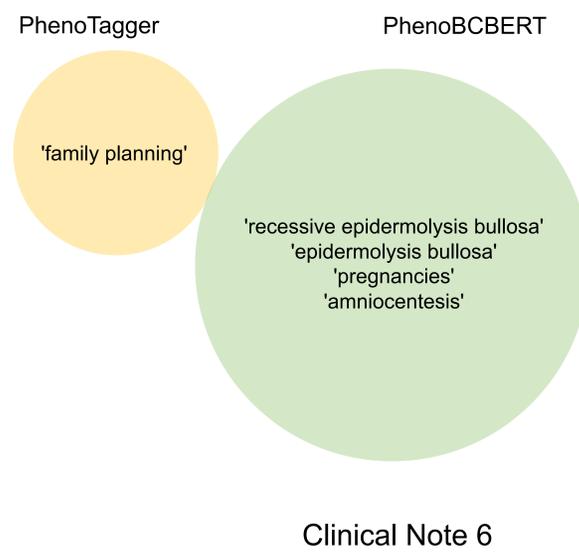

Clinical Note 6

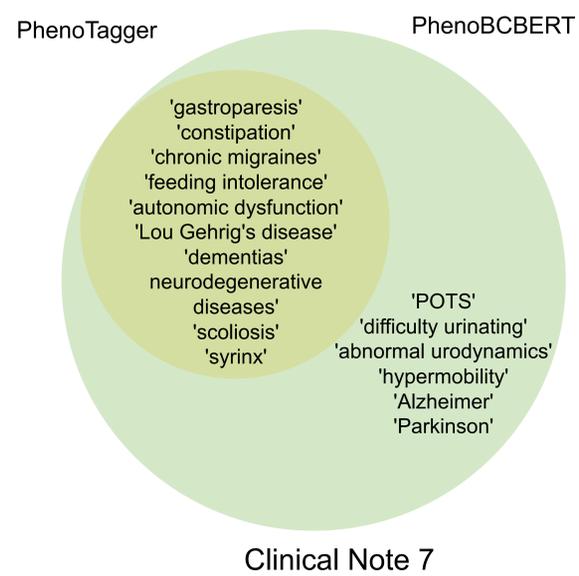

Clinical Note 7

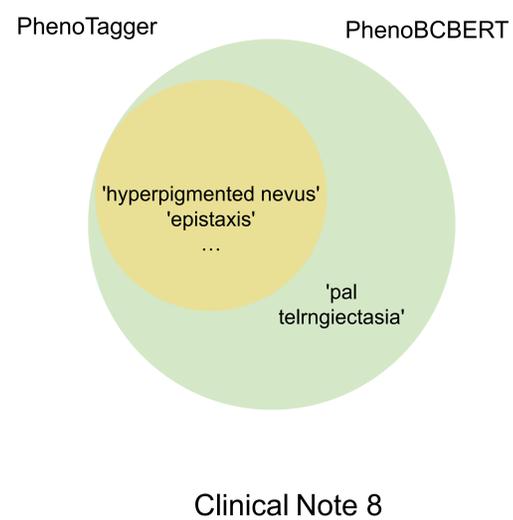

Clinical Note 8

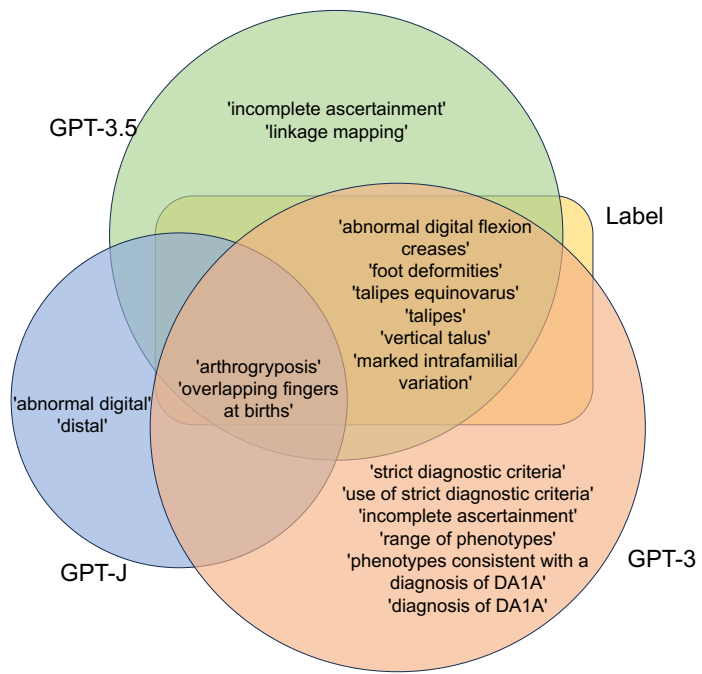

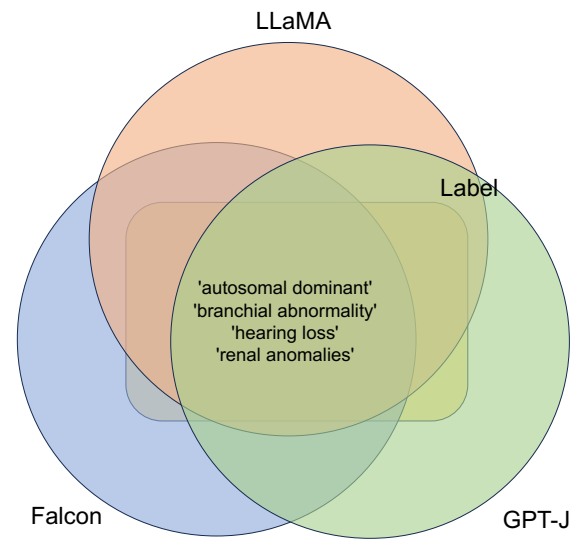

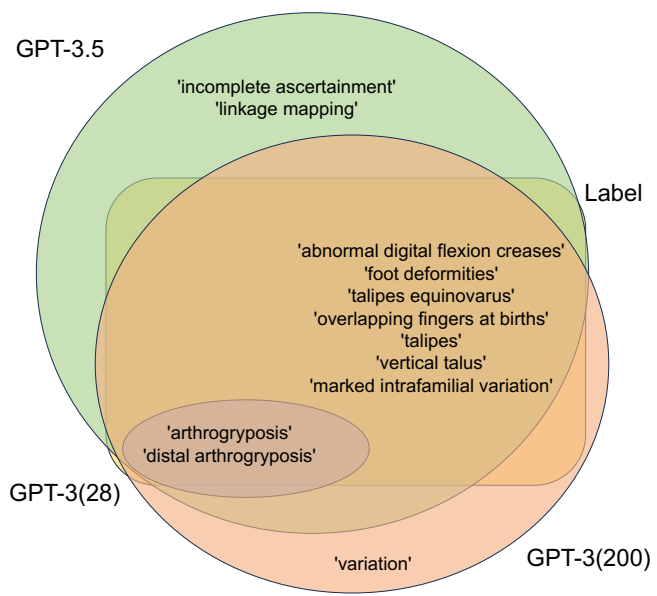

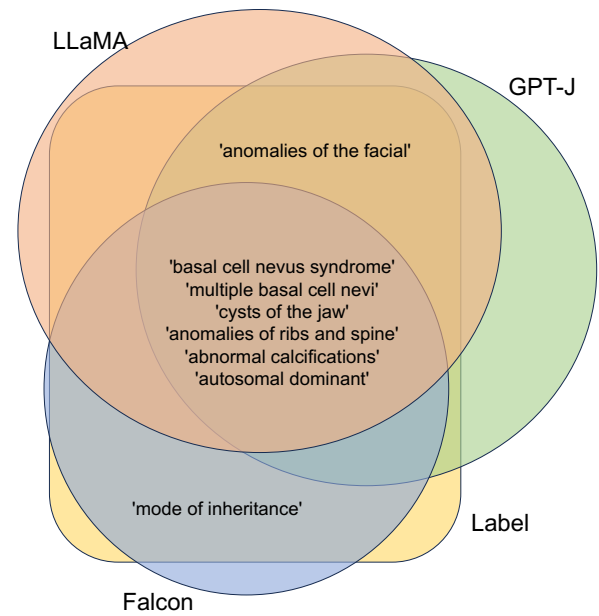

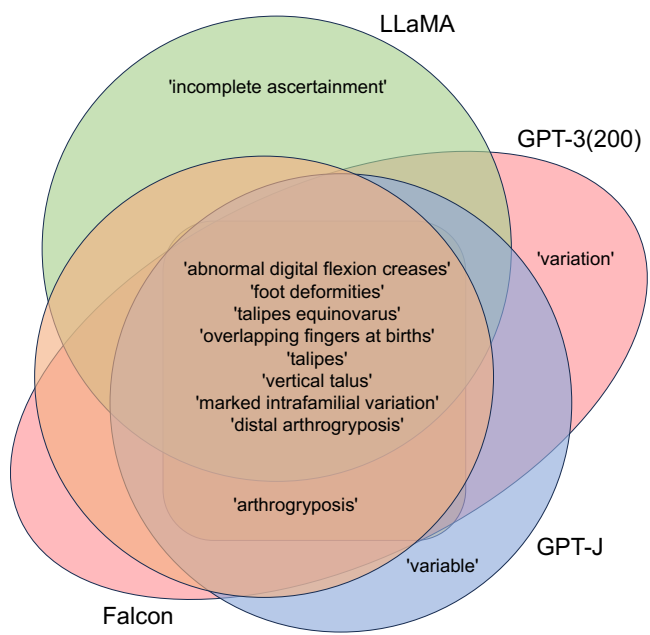

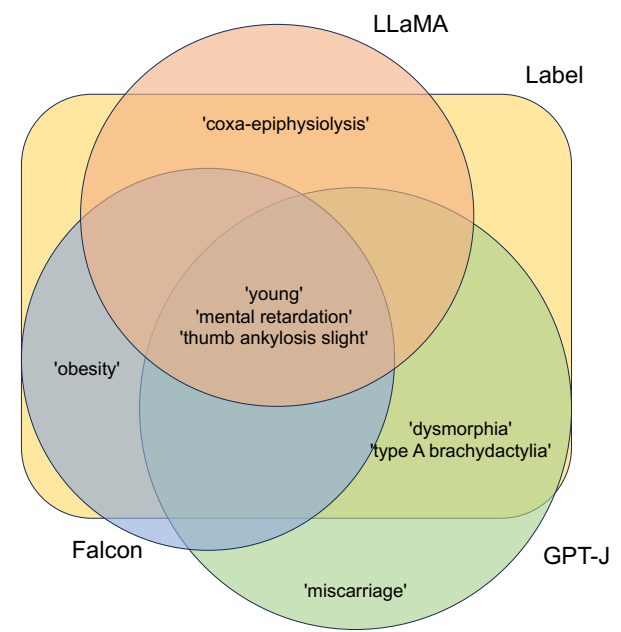

A. Individual 1 is a three-year-old female with developmental delays in expressive speech. Prenatal ultrasound indicated a restricted cerebellar growth and fetal magnetic resonance imaging (MRI) showed ventriculomegaly. A small cerebellum was also reported, but was not seen on review of the images. Prenatal triple screen, nuchal translucency and non-invasive prenatal testing (Harmony) did not indicate chromosomal abnormalities. Labor was noticeable for prolonged rupture of membranes. After birth, she had hyperbilirubinemia, which resolved with sunlight. She had mild muscular hypotonia. She was breastfed, but had problems with latching and swallowing.

C. The first symptoms started at 8 years of age when a severe neurosensory hyopoacusia was found. At 9 years, he developed gait difficulty with stiffness. At the same age, he started to present a dorsal right-convex scoliosis and a retinopathy was detected. Neurological examinations at 16 years showed muscle weakness and diffuse muscle hypotrophy, more severe in limbs and distally. Pyramidal signs such as high tendon reflexes (Babinski and Achilles tendon clonus) were detected and muscle tone was high in limbs. Pes cavus was observed bilaterally, although more evident in the right foot. Achilles tendon retraction was present. The individual presented with distal tremor.

B. Two siblings were referred to the genetics clinic for mild global developmental delay. Individual 1 was born at term following vacuum extraction. Biometry and birth parameters were normal. Pregnancy was complicated as a result of gestational diabetes and the finding of a single umbilical artery on fetal ultrasound. Head circumference was normal, but height and weight were above the 95th centile as of age 18 and 24 months and beyond, respectively. Renal ultrasound revealed a pyelocaliceal dilatation, which regressed spontaneously. Evaluation at age 7 years showed normal intellectual ability, specific learning difficulties mainly related to expressive language, attention deficit hyperactivity disorder, poor fine motor skills, and mild non-specific dysmorphisms including high forehead, arched eyebrows, short columella, micrognatia, and small ears.

PhenoTagger ———
PhenoBCBERT ———
PhenoGPT ———
Negation
Misspelling